\def\beq{\begin{equation}}
\def\seq{\end{equation}}
\def\beqs{\begin{eqnarray}}
\def\seqs{\end{eqnarray}}

\def\D{\Delta_0}
\def\w{\omega}

\def\kp{\kappa_{xx}}
\def\kph{\kappa_{xx}(H)}
\documentstyle[aps,prl,epsf,multicol]{revtex}
\begin{document}
\twocolumn[\hsize\textwidth\columnwidth\hsize\csname@twocolumnfalse\endcsname
\title{Quasiparticle
	thermal conductivity in the vortex state of
	high-T$_c$ cuprates.}
\author{I. Vekhter$^a$ and  A. Houghton$^b$}
\address{${}^a$Department of Physics, University of Guelph, Guelph, 
	Ontario N1G 2W1, Canada\\
	${}^b$Department of Physics, Brown University,
	Providence, RI 02912-1843, USA}
\date{\today}
\maketitle
\tighten
\begin{abstract}
We present the results of a microscopic calculation of the 
longitudinal thermal conductivity, $\kp$, of a $d$-wave superconductor
in the mixed state. Our results show an increase in the thermal
conductivity with the applied field at low temperatures, and a decrease
followed by a nearly field independent $\kph$ at higher temperatures,
in qualitative agreement with the experimental results. We discuss the 
relationship between the slope of the superconducting gap and the plateau
in $\kph$.
\end{abstract}
\pacs{74.25.Fy,74.72.-h, 74.60.-w}
\vspace*{-10pt}
\twocolumn
\vskip.2pc]
\narrowtext

While the origin of the unusual normal state properties of the 
high-T$_c$ cuprates remains controversial, many experimental results 
lend support to the belief that the superconducting state
of these materials is relatively mundane, and consists of a pair
condensate with an ``unconventional'' order parameter, believed
to be of  $d$-wave symmetry, and
well-defined quasiparticle excitations above it.
However, transport measurements in the vortex state
yield some
surprising results which question whether 
the low-temperature physics of the cuprates is
well understood.

Perhaps the  most notable such example is the
observation of a plateau-like feature in the magnetic field dependence
of the longitudinal thermal conductivity, $\kph$,  in
superconducting Bi$_2$Sr$_2$CaCu$_2$O$_{8+\delta}$ (BSCCO)
\cite{ong1} and
YBa$_2$Cu$_3$O$_{7-\delta}$ (YBCO) \cite{ong2}. 
At temperatures above a few Kelvin
 $\kp(H)$ initially decreases 
with applied magnetic field, $H$,
before becoming 
nearly field independent.\cite{ong1,ong2,kamran1,ando} 
The kink and hysteresis in  $\kp(H)$ 
observed in zero-field cooled BSCCO \cite{ong1,kamran1}
have not been found in YBCO \cite{ong2} or in the field-cooled BSCCO samples
\cite{ando}, and we take the point of view that they are not a generic feature 
of the $\kp(H)$ dependence in cuprates.
In some samples the ``plateau'' is not reached, and 
$\kp(H)$ decreases 
continuously for $H\leq 14$ Tesla,\cite{ong2,ando}
as it does for conventional superconductors
for $H\ll H_{c2}$.\cite{sousa}
This is in sharp contrast to the behavior of the thermal conductivity 
of the cuprates 
at ultra-low temperatures, which increases with $H$.\cite{kamran1,may}
Moreover,  both features 
have been observed in the same sample\cite{kamran1}, suggesting
that the two regimes should be understood within the same approach, which
presents
a challenge to theories of quasiparticle transport in the vortex state
of the cuprates.

Existing theories are based on the observation
that, for a single vortex, the properties of 
superconductors with lines of nodes in the energy gap
are dominated by
extended quasiparticle states, rather than,
as in the $s$-wave case, by localized states
in the vortex cores.\cite{volovik} 
The magnetic field is taken into account semiclassically, by introducing
a Doppler energy shift due to 
the circulating supercurrents.\cite{volovik}
In one approach 
physical quantities are computed locally and 
are averaged over a unit cell
of the vortex lattice\cite{kubert1}.
The results obtained within this framework
describe experimental results quite well\cite{kubert1,kubert2,vekhter}, 
and 
the predictions for $\kph$
by K\"ubert and Hirschfeld\cite{kubert2} 
fit the ultralow-$T$ measurements.\cite{may} 
The results of Ref.\cite{kubert2} show
 a decrease in $\kph$ at higher $T\ll T_c$, 
but not a plateau.
A similar approach was used
for heavy fermion superconductors.\cite{barash}

Franz\cite{franz} has 
pointed out  that the averaging procedure used in 
Refs.\cite{kubert1,kubert2,vekhter}
neglected  
scattering from the vortex lattice. He included the scattering
semi-phenomenologically, assuming a completely disordered 
vortex lattice and a Matthiesen type rule for the impurity and vortex
contributions to the lifetime\cite{leggett}.
 He predicted
that at any $T\le T_c$  $\kph$ will approach asymptotically the
universal value \cite{lee,graf} $\kappa_{00}=\pi T N_0 v^2/6\D$,
where $N_0$ is the normal state density of states at
the Fermi surface, $v$ is the Fermi velocity, and $\D$ is the gap amplitude. 
His results are in qualitative agreement with the results of
Ref.\cite{ong2}, but the theory requires additional assumptions 
to explain
the low-$T$ limit. 
Consequently, no existing theory explains 
{ both} the low-$T$ results and the plateau.

In this work
we present a microscopic theory of the thermal conductivity of 
a two-dimensional clean ($l\gg \xi_0$, where $l$ is the mean free path, and
$\xi_0$ is the coherence length)
 $d$-wave superconductor in the vortex state, with the field
perpendicular to the basal plane, and thermal current in the plane.
Our results
fit the low-$T$ data of Ref.\cite{may} and show 
a plateau-like feature at higher $T$. 
We employ the quasiclassical method
\cite{E,LO}, in which the basic quantity is the matrix
Green's function
integrated over the quasiparticle energy,
$$\widehat g(\phi,{\bf R}; \omega_n,\omega_{n^{\prime}})=
    \int { d \zeta_p \over \pi}
                \widehat 
                G(p,{\bf R}; \omega_n,\omega_{n^{\prime}})
= 
              \pmatrix{g& -f\cr
                                   f^{\dagger} & \bar{g}\cr}.
$$
Here the diagonal elements, $g$ and $\bar g$, are the particle and hole
propagators, respectively, the off-diagonal elements, 
Gorkov's anomalous functions $f$ and $f^{\dagger}$, are
probability amplitudes for the creation or destruction of a Cooper pair,
{\bf R} is the center of mass coordinate,  
$\zeta_p$ is the energy of a quasiparticle with momentum $p$ measured
relative to the Fermi surface, and $\phi$ is the angle parameterizing the
cylindrical Fermi surface. 
The spatial dependence
of the vortex lattice (VL) is modeled by an Abrikosov-like solution
$\Delta({\bf R},\phi)=\Delta({\bf R})\varphi(\phi)$, where
$\Delta({\bf R})=\sum_{k_y} C_{k_y} e^{i k_y y} 
\exp
(-(x - \Lambda^2 k_y)^2
                          / 2 \Lambda^2)$,
the magnetic length $\Lambda=(2eH)^{-1/2}$ ( 
$\hbar=c=1$ throughout the paper), 
and $\varphi(\phi)=\cos 2\phi$ is the angular dependence of the gap.
\cite{vlattice}

We use the quasiclassical method in the framework proposed in Ref.\cite{BPT}
for $s$-wave superconductors near $H_{c2}$.
 In that
approach the diagonal elements of $\widehat g$ are approximated by 
their spatial averages, while the exact spatial 
dependence of the off-diagonal elements is kept.
The authors of Ref.\cite{BPT} noticed
that $g$ and $\bar g$ are periodic in {\bf R} with the periodicity of
the vortex lattice, and that 
Fourier components $g_{\bf K}$ with reciprocal vortex lattice 
vectors ${\bf K}\neq 0$ are smaller than $g_{{\bf K}=0}$ 
(which is the spatial average of $g$) by a factor
$\exp( - \Lambda^2 K^2)$.  
This approach provides a very accurate description of the
vortex state in conventional superconductors.
\cite{BPT,houghton,purvis,pesch,klimesch}
The analysis of Ref.\cite{BPT} relies on the fact 
that the spatial dependence of the order parameter
is given by an Abrikosov-like solution of the linearized Ginzburg-Landau
equations, and therefore works especially well 
in extreme type-II superconductors where it can be extended over almost 
the entire region of linear magnetization.\cite{ehbrandt} 
For an $s$-wave superconductor it breaks down
at fields where  
properties are determined by the states bound to vortex cores.
In unconventional superconductors where 
properties are dominated by extended
states\cite{volovik} the approach works
down to fields close to $H_{c1}$, 
i.e. for all experimentally relevant fields.\cite{makiwon} 

The leading order quasiclassical propagator is determined
from equations \cite{pesch,vekhter1}
\begin{eqnarray}
\label{eqf0}
(2 \widetilde\omega_n  + {\bf v}( \nabla - 2 i e {\cal A})) f&=&
2 i \Delta({\bf R}) g \cos 2\phi
\\
\label{eqfd}
(2 \widetilde\omega_n  - {\bf v}( \nabla + 2 i e {\cal A})) f^{\dagger}&=&
2 i \Delta^{\star}({\bf R})  g\cos 2\phi,
\end{eqnarray}
complemented by the
normalization condition $g^2-ff^\dagger=-1$.
Here ${\cal A}=(0,Hx,0)$, and the renormalized frequency,
$i\widetilde\omega_n=i\omega_n+i\sigma(i\widetilde\omega_n)$,
depends on the self-energy
due to impurity scattering, $\sigma$, which  is determined within
the $t$-matrix approximation\cite{hirschfeld}. 
We ignore 
corrections to the impurity vertex due to the magnetic field,
which is justified when $\Lambda\ll l$. Since 
$\Lambda\simeq 183{\rm \AA}/\sqrt{H,{\rm Tesla}}$,
for clean samples this condition is satisfied at least above 
0.1-0.5 T.

The solution of Eqs.(\ref{eqf0}),(\ref{eqfd}) is known to be
\cite{pesch,vekhter1}
$g=- {i }{\rm sgn}(\omega_n ) 
P(\phi,\widetilde\omega_n)$, where
$ P(\phi,\widetilde\omega_n)=
[ 1- i \sqrt{\pi} 
                      \bigl({2 \Lambda \Delta /v }\bigr)^2
                           W^{\prime}
                    \bigl (u_n)\cos^2 2\phi]^{-1/2}$,
the function $W(u)=e^{-u^2}{\rm erfc}(-iu)$, 
$u_n= {2 i \widetilde\omega_n \Lambda{\rm sgn}
                                     (\omega_n)
                                     /v}$,
and $\Delta$ is the spatial average of
the amplitude of the
order parameter. 
In the limit $H=0$ setting $\Delta$ equal to the gap amplitude, $\D$,
we recover the BCS Green's function and
consequently all the results of the 
standard ``dirty $d$-wave'' theory.\cite{goldenfeld}
 For a pure sample at
low fields the residual density of states (DOS) 
$N(0)/N_0=(v/\Lambda\Delta_0\pi\sqrt 2)
		\ln(8\sqrt 2\Lambda\Delta_0/v)
\propto\sqrt H\ln(H_{c2}/H)$;
the field dependence 
differs
by a logarithm from the single vortex result of Volovik\cite{volovik}.
For the range of fields where comparison with experiment has
been made the logarithm is nearly constant,
and the result for the DOS agrees with that of 
of Ref.\cite{kubert1} up to a numerical factor.
In the following
we consider scattering in the unitarity and Born limits,
so that the parameters of the theory are
the normal state scattering rate, $\Gamma$, and the dimensionless quantity
$\Lambda \Delta /v$. 
The latter can be estimated by
setting $\Delta\simeq \D(1-H/H_{c2})^{1/2}$, 
and noticing that $\Lambda \D/v=(\Lambda k_f/2)(v_2/v)$,
where $v_2$ is the slope of the gap at the node.
In BSCCO $k_f\simeq 0.737$ \AA$^{-1}$, 
and is relatively doping independent,\cite{norman1}
then $\Lambda \Delta /v\sim \alpha \sqrt{(1-H/H_{c2})/H}$;
where, for field in Tesla,
$\alpha\sim 3.5$ for the overdoped case, where $v/v_2\sim 20$,
but becomes smaller on the underdoped side.\cite{norman}
In YBCO, taking $k_f\sim 0.8$\AA$^{-1}$\cite{schnabel})
we obtain
$\alpha\sim 5$ if $v/v_2\sim 14$\cite{may}, or twice that value
if $v/v_2\sim 7$.\cite{leewen}

The heat current can now be determined from the linear response
equations \cite{klimesch,vekhter1}
\beqs
\label{jh}
{\bf j}_h&=&\pi N_0 \lim_{i\omega_0\rightarrow i0^+}
\sum_{\omega_n}\int d\phi {\bf v}(\phi) ({\bf v}(\phi)\cdot\nabla T)
\\
\nonumber
	&&\times	\omega_n(\omega_n-\omega_0)
{g(\omega_n)-g(\omega_n-\omega_0)\over i\widetilde\omega_0 +2b},
\seqs
where
$b=(\Delta f^\dagger(\omega_n) +\Delta^\star f(\omega_n-\omega_0))
	/ (g(\omega_n)-g(\omega_n-\omega_0))$.
This leads to the thermal conductivity 
\begin{eqnarray}
\label{kpa}
&&{\kp\over T}={N_0 v^2 \over 4\pi}
\int_0^\infty {d\omega\over T} \biggl({\omega\over T}\biggr)^2
\cosh^{-2}\biggl({\omega\over 2T}\biggr) 
\\
\nonumber
&&\qquad\qquad\times
\int_0^{2\pi}d\phi
\cos^2\phi
 \biggl[Re P(\widetilde\omega) \biggr]\tau(\widetilde\omega,\phi),
\\
\label{kernel}
&&{1\over2\tau(\widetilde\omega,\phi)}=
Re\sigma(\widetilde\omega)
\\
\nonumber
&& \qquad\qquad +
2\sqrt\pi{\Lambda\Delta^2\over v} 
{Re [P(\widetilde\omega)W(u)]
\over Re P(\widetilde\omega)}\cos^2 2\phi.
\end{eqnarray}
Since $Re [P(\widetilde\omega, \phi)]$ is the 
angular dependent DOS, and  $\tau(\widetilde\omega,\phi)$ is a transport 
lifetime, 
Eqs.(\ref{kpa}) and (\ref{kernel}) generalize the normal state
expression
$\kp\equiv\kappa_n=\pi^2 N_0 v^2 T\tau/6$.
There is a 
contribution to the quasiparticle lifetime from Andreev
scattering off the vortex lattice,
which has the same symmetry on the Fermi surface as the gap function, and is
therefore the more important the higher the quasiparticle energy.
This contribution appears
even for a
periodic vortex lattice, 
since Bloch's theorem only
forbids scattering from a periodic potential {\it in the absence} 
of impurity
broadening of the quasiparticle wave-packet, not in
the presence of impurities. 
Note
that the dependence of the DOS and the scattering rate
on angle cannot be disentangled, 
in contrast to the assumption
of Ref.\cite{franz}. Also, it is clear from the form of Eq.(\ref{kpa})
that a simple scaling form as a function of $T/\sqrt H$ cannot be expected.

First, we consider the limit $T=0$, where analytic results can be
obtained. It is convenient to analyze
the thermal conductivity in terms of a transport lifetime
$\tau_H$, defining
$\kph/T =\pi^2 N_0 v^2 \tau_H/6$, and expressing $\tau_H$ in terms
of the lifetime $\tau_{eff}\equiv 1/(2\gamma)$, where
$\gamma=\sigma(\w=0)$ is the zero-frequency scattering rate.
For $H=0$ we obtain
$\tau_H=k E(k)/\pi\D$, where
$E$ is the complete elliptic integral, and
$k=(1+(2\Delta_0\tau_{eff})^{-2})^{-1/2}$
in agreement with Refs.\cite{sun,graf}. 
For $\D\tau_{eff}\gg 1$ we recover the
universal limit $\kp=\kappa_{00}\equiv\pi T N_0 v^2/6\D$.\cite{lee,graf}
At finite fields the integral in 
Eq.(\ref{kpa}) is evaluated approximately to give
$\tau_H\simeq(\tau_{eff}v/4\sqrt\pi\Lambda\D^2)^{1/2}$. In the ``clean''
limit
$v/\Lambda\gg \gamma$ 
for the important cases of Born ($\Gamma_B$) and unitarity 
($\Gamma_U$) scattering,
we find
\begin{equation}
\label{tauHex}
\tau_H\simeq\cases{({\pi\over 32})^{1/4} (\Gamma_B\Delta_0)^{-1/2}
	\ln^{-1/2}({8\sqrt 2\Lambda\Delta_0\over v});\cr
	(128\pi^3)^{-1/4}(\Gamma_U\Delta_0)^{-1/2}({v\over\Lambda\Delta_0})
	\ln^{1/2} ({8\sqrt 2\Lambda\Delta_0\over v}).
\cr}
\end{equation}
This expression is in a good agreement with the numerical results at
intermediate fields.  In Fig. 1
we use the numerical evaluation of Eq. (\ref{kpa})
to compare $\kph$ for Born and unitarity scattering.
In the case of Born scattering  $\kph$ only reaches the universal limit
at exponentially small fields, and
is 
approximately field-independent at higher $H$, Eq.(\ref{tauHex}). 
In the unitarity limit, on the
other hand, $\kph$ increases approximately as $\sqrt H$ at low fields
and the approach to $\kappa_{00}$ is clearly seen.
At low $T$ it is seen that $\kph$ tends to $\kappa_{00}$
over a field range of order a few Tesla\cite{may},
and that the field dependence is close to $\sqrt H$ at low $T$.\cite{kamran1}
This suggests
that 
impurity scattering is
close to the unitarity limit at low temperatures, and 
we consider this limit hereafter.
The $H$-dependence of $\tau_H$ differs from that found in the  Doppler
shift theory\cite{kubert2}, 
however, since $\sqrt{\ln H}$ varies slowly, for 
experimentally accessible
fields the behavior of $\kph$
is close in to that obtained by 
K\"ubert and Hirschfeld.\cite{kubert2}
In Fig. 2 we fit
the data of Ref.\cite{may}.  Using the  
the value of $v/v_2=14$ from Refs.\cite{may,louis} we find
$\Gamma$ to be about  
4 times greater
than the estimate of Ref.\cite{may}.
In the present theory 
$\kph$ increases faster with field the smaller
the slope of the gap,
approximately as $\sqrt{H/\alpha^2\Gamma}$, see Eq. (\ref{tauHex}),
so that the same data can be fit on assuming a smaller impurity
concentration if we allow for a larger ratio $v_2/v$.\cite{leewen}
This also implies that a smaller slope of the gap can result in a larger 
increase in $\kph$ for a dirtier sample, see Fig. 2, which 
may be a part of the explanation of 
a faster increase of $\kph$ in nominally dirtier
BSCCO compared to YBCO.\cite{kamran1}

At finite temperatures the integration in Eq.(\ref{kpa}) has to
be carried out numerically; 
we confine ourselves to the regime where the gap is temperature 
independent. 
We have checked and find agreement
with previous results at $H=0$.\cite{graf}
$\kappa(0)/T$ 
increases rapidly for $T\gg \gamma$, and for clean samples at $T\sim 0.1\D$
it exceeds
$\kappa_{00}$ by two orders of magnitude. 
At the same time
at strong fields,
$v/\Lambda\gg T$, $\kph/T$ is controlled
by the field
strength and is only weakly temperature dependent.
The field dependence at $T=0$ is weak,
close to $\sqrt H$, so that 
for $H\sim 10$ T $\kph/T\leq 10\kappa_{00}/T$.
 An immediate conclusion
is that for $T >\gamma$ 
there is a steep drop in $\kph$ at low fields,
followed by a crossover to the $\sqrt H$ like behavior.
This crossover occurs when $v/\Lambda\sim T$, at a field
$H^*\propto T^2$, as indicated experimentally.\cite{ong1,kamran1}

In  Fig.3 we show the results:
at high $T$  the ratio $\kph/\kp(0)$ decreases 
continuously as a function of field,
however as the temperature is lowered a plateau 
develops, and at even lower $T$ a pronounced minimum appears
in the field dependence. 
As the lower panel demonstrates there is a transition
to an increasing $\kph$ at the lowest temperatures.
Numerically, we find that 
at fixed $T$ a smaller value of 
$\alpha$ corresponds to a more pronounced plateau
in $\kph$. This implies  that the nearly field independent
thermal conductivity should be more evident
in the underdoped cuprates.\cite{norman}

To summarize, we give a theory of the thermal conductivity of a $d$-wave
superconductor in the vortex state as a function of field at  $T\ll T_c$. 
In zero field and at low $T$ we recover the universal 
results\cite{lee,graf}. We find that there exists a contribution to
the transport lifetime due to scattering off the vortices, which has the
same symmetry as the gap function. Consequently, at low temperatures when 
most of the quasiparticles are near the gap nodes it plays a minor role and
the effect of the field is to increase the DOS and the thermal 
conductivity. In contrast, at high $T$ the main effect of the vortices
is to introduce a new scattering mechanism, and $\kph$ decreases sharply to a
plateau-like feature at higher fields (a broad minimum); 
the crossover field scales with $T^2$.
This behavior is in agreement with experimental results.
\cite{ong1,ong2,kamran1,ando,may}.
Finally, it is important
to compare the predictions of our theory with the measurements 
of the transverse (Hall) thermal conductivity, $\kappa_{xy}$, \cite{ong2}
to which phonons do not contribute. At low fields $\kappa_{xy}$ is given by
Eq.(\ref{kpa}) with the kernel 
$\tau_{xy}(\widetilde\omega,\phi)=
\w_c\tau(\widetilde\omega,\phi)/2Re[\sigma(\widetilde\omega)]$.
However, as the measurements are taken at $T\ge 20$K, 
in a regime where it is necessary to include 
inelastic scattering in the analysis of both longitudinal and
transverse thermal conductivity,\cite{putikka} we defer a
discussion of this
issue until later. 

We are grateful to M. Chiao, M. Franz, D. Maslov, and
especially P. J. Hirschfeld and 
L. Taillefer for discussions, and to K. Behnia for 
correspondence. A.H. acknowledges the hospitality of ITP Santa Barbara, 
where part of this work was done.

\begin{figure}
\label{fig1}
\epsfxsize=2.5in
\epsfbox{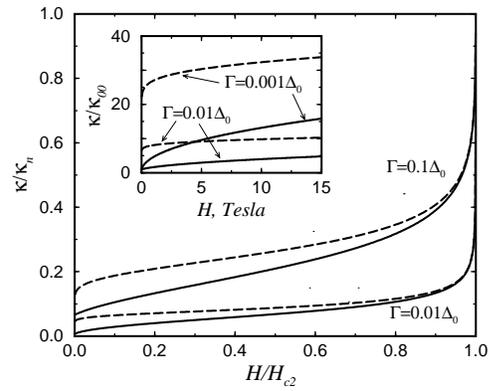}
\caption{\narrowtext
Field dependence of the thermal conductivity at $T=0$ for scattering
in the unitarity (solid lines) and
Born (dashed lines) limits. Inset: low field behavior, normalized by
the universal value $\kappa_{00}$. $H_{c2}=150$ T.}
\end{figure}
\begin{figure}
\label{fig2}
\epsfxsize=2.7in
\epsfbox{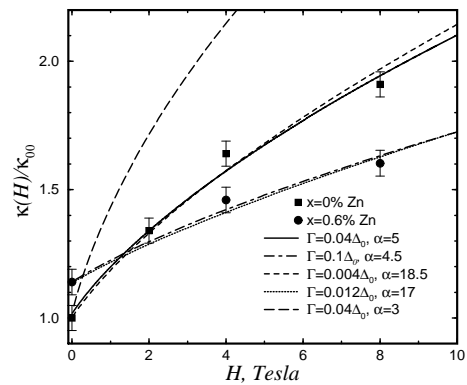}
\caption{\narrowtext Field dependence of $\kph$ at $T=0$ 
compared to data from Ref.[6].}
\end{figure}
\begin{figure}
\label{fig3}
\epsfxsize=2.7in
\epsfbox{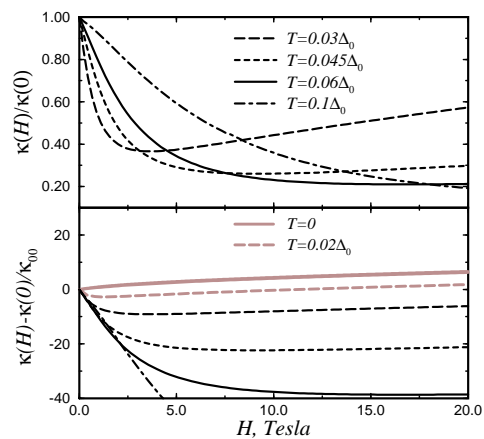}
\caption{\narrowtext
Top: field dependence of $\kph/\kappa(0)$ for $\Gamma=0.003\D$
and $\alpha=7$ at temperatures where $\kph$ is decreasing; bottom:
the {\it change} in the thermal conductivity as a function of field for
different temperatures.}
\end{figure}

\end{document}